\documentclass[rnote]{aa} % for the research notes
\usepackage{graphicx}
\usepackage{txfonts}
\begin{document}
   \title{Additional spectra of asteroid 1996 FG3, backup target of the ESA MarcoPolo-R mission}

%   \subtitle{I. Overviewing the $\kappa$-mechanism}

   \author{J. de Le\'on\inst{1,3}
          \and
          V. Lorenzi\inst{2}
          \and
          V. Al\'i-Lagoa\inst{3,4}
          \and
          J. Licandro\inst{3,4}
          \and
          N. Pinilla-Alonso\inst{5}
          \and
          H. Campins\inst{6}
          }

   \institute{Departamento de Edafolog\'ia y Geolog\'ia, Universidad de La Laguna (ULL)
              Avda. Astrof\'isico Francisco S\'anchez, s/n, E-38205, La Laguna, Tenerife, Spain\\
              \email{julia.de.leon@ull.es}\\
         \and
             Fundaci\'on Galileo Galilei - INAF, Rambla Jos\'e Ana Fern\'andez P\'erez 7
             37812, La Palma, Spain\\
         \and
             Instituto de Astrof\'isica de Canarias (IAC)
             C/V\'ia L\'actea s/n, E-38205 La Laguna, Spain\\
         \and
             Departamento de Astrof\'{\i}sica, Universidad de La Laguna (ULL), 
	           E-38205 La Laguna, Spain\\
	       \and
	           Earth and Planetary Sciences Department, University of Tennessee, Knoxville, TN 37996, USA\\
	       \and
	           Department of Physics, University of Central Florida, 
             PO Box 162385, Orlando, FL 32816.2385\\
             }

   \date{Received March 20, 2013; accepted June 25, 2013}

% \abstract{}{}{}{}{} 
% 5 {} token are mandatory
 
  \abstract
  % context heading (optional)
  % {} leave it empty if necessary  
   {Near-Earth binary asteroid (175706) 1996 FG$_3$ is the current backup target of the ESA MarcoPolo-R mission, selected for the study phase of ESA M3 missions. It is a primitive (C-type) asteroid that shows significant variation in its visible and near-infrared spectra.}
  % aims heading (mandatory)
   {Here we present new visible and near-infrared spectra of 1996 FG$_3$. We compare our new data with other published spectra, analysing the variation in the spectral slope. The asteroid will not be observable again over the next three years at least.}
  % methods heading (mandatory)
   {We obtained visible and near-infrared spectra using DOLORES and NICS instruments, respectively, at the Telescopio Nazionale Galileo (TNG), a 3.6m telescope located at El Roque de los Muchachos Observatory in La Palma, Spain. To compare with other published spectra of the asteroid, we computed the spectral slope $S'$, and studied any plausible correlation of this quantity with the phase angle ($\alpha$).}
  % results heading (mandatory)
   {In the case of visible spectra, we find a variation in spectral slope of $\Delta$$S'$ = 0.15 $\pm$ 0.10 \%/10$^3$\AA / $^{\circ}$ for 3$^{\circ}$ $<$ $\alpha$ $<$ 18$^{\circ}$, which is in good agreement with the values found in the literature for the phase reddening effect. In the case of the near-infrared, there seems to be a trend between the reddening of the spectra and the phase angle, excluding one point. We find a variation in the slope of $\Delta$$S' = 0.04 \pm 0.08  \%/10^3\AA / ^{\circ}$ for $6^{\circ} < \alpha < 51^{\circ}$. Our computed variation in $S'$ is in good agreement with the only two values found in the literature for the phase reddening in the near-infrared.}
  % conclusions heading (optional), leave it empty if necessary 
   {The variation in the spectral slope of asteroid 1996 FG$_3$ shows a trend with the phase angle at the time of the observations, both in the visible and the near-infrared. It is worth noting that, to fully explain this spectral variability we should take into account other factors, like the position of the secondary component of the binary asteroid 1999 FG$_3$ with respect to the primary, or the spin axis orientation at the time of the observations. More data are necessary for an analysis of this kind.}

   \keywords{minor planets, asteroids: individual: 1996 FG$_3$ -- methods: observational -- techniques: spectroscopic
               }  
\authorrunning{de Le\'on et al.}
\titlerunning{Additional spectra of asteroid 1996 FG$_3$}					
\maketitle
%
%________________________________________________________________

\section{Introduction}

Binary asteroid (175706) 1996 FG$_3$ (hereafter FG3) is currently the backup target for the ESA MarcoPolo-R mission, selected for the assessment study phase of ESA M3 missions. This is a near-Earth binary system with semimajor axis $a$ = 1.054 AU, eccentricity $e$ = 0.35, and inclination $i = 1.98^{\circ}$, and a mutual orbital period of P = 16.135 $\pm$ 0.005 h (Scheirich and Pravec \cite{scheirich09}). The primary component has a diameter of about 1.40--1.83 km and has a fast spin rate (3.6 h). The secondary orbits the primary with $e$ = 0.05 $\pm$ 0.05 and $a \sim 1.4$ times the primary's radius and has an estimated diameter of 0.43--0.51 km (Pravec et al. \cite{pravec00}; Mottola and Lahulla \cite{mottola00}; Walsh et al. \cite{walsh12}). 

The most recent albedo determination from thermal infrared observations gives a value of $p_V = 0.039 \pm 0.012$ (Walsh et al. \cite{walsh12}), which is consistent with the asteroid being taxonomically classified as a C-type object. The best meteorite analogs are CM2 (de Le\'on et al. \cite{deleon11}; Popescu et al. \cite{popescu12}) and CV3 (Rivkin et al. \cite{rivkin12}) carbonaceous chondrites. This primitive composition makes this asteroid a particularly interesting target for a space mission. Primitive asteroids are believed to consist of carbon-rich and organic materials that have not been altered by processes such as melting and mixing that occurred during the early stages of the formation of the solar system. In addition, the binary nature of the target will allow more precise measurements of mass, gravity, and density than a single object does, and it will offer additional insights into the geology and geophysics of the system. 

Although the primitive nature of FG3 is not questioned, spectra obtained by different authors during its close approach between late 2011 and early 2012 show a significant variation in spectral slope. Here we present additional visible and near-infrared spectra of FG3, which are the latest observations obtained of this object. This particular asteroid will not be observable again for at least the next three years, either because it is too faint or because it will not be visible. 

\section{Observations and data reduction}\label{obs}

\begin{figure}
   \centering
   \includegraphics[width=8cm]{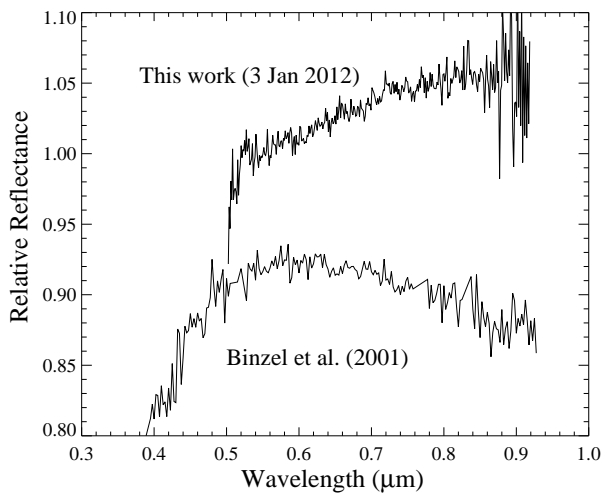}
      \caption{Visible spectra of asteroid FG3. Spectra are normalized to unity at 0.55 $\mu$m. The spectrum presented in this work is compared with visible spectrum from Binzel et al. (\cite{binzel01}), which has been vertically offset for clarity.}\label{figure1}
   \end{figure}

The visible spectrum of FG3 was obtained on January 3, 2012, with the 3.6m Telescopio Nazionale Galileo (TNG) using the DOLORES spectrograph. The low resolution red (LR-R) grism was used covering the 0.50--0.95 $\mu$m spectral range with a resolution of 2.61\AA/pix. The object was centered in a 2'' slit, which was oriented to the parallactic angle to minimize losses due to atmospheric dispersion. Three spectra of 600 seconds of integration time each were obtained, shifting the object 10'' in the slit direction between consecutive spectra to better correct the fringing. Images were bias and flat-field corrected using standard procedures. The two-dimensional spectra were extracted, sky background subtracted, and collapsed to one dimension. Wavelength calibration was done using Ne and Hg lamps. The three individual spectra of the asteroid, obtained at different positions in the slit, were averaged. To obtain the  asteroid's reflectance spectrum, we observed two solar analog stars from the Landolt catalogue (Landolt \cite{landolt92}) with similar airmass to that of the asteroid: SA 98-978 and SA 102-1081. The asteroid spectrum was divided by the individual spectrum of each solar analog, and the resulting spectra were finally averaged and normalized to unity at 0.55 $\mu$m. Our final spectrum is shown in Fig. \ref{figure1}, together with that from Binzel et al. (\cite{binzel01}), obtained on January 26, 1998 and is the only visible spectrum of FG3 published up to now.

Low resolution near-infrared spectra of FG3 were also obtained with the TNG on December 21 and 24, 2011, using the low resolution mode of Near Infrared Camera Spectrograph (NICS) and the Amici prism disperser, which covers the 0.8--2.5 $\mu$m spectral range. Two different slits were used on the first and second nights, 2.0'' and 1.0'', respectively. The slit was in both cases oriented along the parallactic angle, and the tracking was performed at the asteroid's proper motion. The acquisition consisted of two series of short exposure images offsetting the object between positions $A$ and $B$ in the slit direction. This process was repeated and a number of $ABBA$ cycles were acquired, with a total on-object exposure time of 1440 seconds. We observed two solar analog stars from the Landolt catalogue: SA 98-978 and SA 115-271. The reduction procedure followed de Le\'on et al. (\cite{deleon10}). After a standard bias and flat field correction, we subtracted consecutive $A$ and $B$ exposures from each $ABBA$ cycle, obtaining individual images from which 1D spectra were extracted and wavelength calibrated. These individual spectra were then averaged and the result was divided by the individual spectrum of each solar analog star. The resulting spectra were finally averaged and normalized to unity at 1.0 $\mu$m (see top panel of Fig. \ref{figure2}). 

   \begin{table}
   \caption{Observational parameters for the FG3's spectra}
   \label{table1}      
   \centering                          
   \begin{tabular}{l c c c c}        
   \hline\hline\\[-3mm]
   Source & Date & $r$  & $\alpha$     & $ S'$\\
          &      & (AU) & ($^{\circ}$) & \%/1000\AA\\ 
   \hline\\[-2mm]
   Visible & & & & \\
   \hline\\[-2mm]
   Binzel [1] & 26/01/1998 & 1.382 & 2.8 & -0.600 $\pm$ 0.200\\
   This work  & 03/01/2012 & 1.222 & 18.0 & 1.620 $\pm$ 0.300 \\
   \hline\\[-2mm]
   Near-infrared & & & & \\
   \hline\\[-2mm]
   Binzel [2] & 30/03/2009 & 1.226 & 8.3 & 0.160 $\pm$ 0.100 \\
   Binzel [2] & 27/04/2009 & 1.083 & 58.5 & -0.469 $\pm$ 0.300 \\
   de Le\'on [3] & 10/01/2011 & 1.354 & 22.5 & 1.694 $\pm$ 0.500 \\
   Binzel [4] & 01/12/2011 & 1.046 & 51.3 & 1.859 $\pm$ 0.050 \\
   Rivkin [5] & 06/12/2011 & 1.075 & 35.7 & 1.165 $\pm$ 0.060 \\
   This work & 21/12/2011 & 1.157 &  5.8 & 0.183 $\pm$ 0.300 \\
   This work & 24/12/2011 & 1.172 & 6.8 & 0.572 $\pm$ 0.100 \\ 
   \hline\\[-2mm]                              
   \end{tabular}
   \tablefoot{[1] Binzel et al. (2001); [2] MIT-UH-IRTF (MINUS, http://smass.mit.edu/minus.html); [3] de Le\'on et al. (2011b); [4] Binzel et al. (2012); [5] Rivkin et al. (2012). Although errors in $S'$ are obtained considering the dispersion of the data points in relative reflectance, it is important to note that they are in fact dominated by the division of the spectra of the solar analog. See text for further details.}
   \end{table}

\section{Spectral analysis}\label{spec}

Comparing the visible and near-infrared spectra of FG3 presented in this work with the spectra previously published one can clearly see a slope variation. In this section we analyze how significant this variation is and present a possible explanation that could account for it.

\subsection{Visible spectra}

As mentioned in the previous section, Fig. \ref{figure1} shows the only two published visible spectra of FG3: the one presented here and the one from Binzel et al. (\cite{binzel01}). 
We compute the spectral slope $S' = (dS/d\lambda)/S_{5500}$ in units of \%/10$^3$\AA \ (Jewitt \& Luu \cite{jewitt90}), in the range between 0.55 and 0.90 $\mu$m. The resulting value for each spectrum is shown in Table \ref{table1}, together with the distance to the Sun ($r$) and the phase angle ($\alpha$) at the time of the observation. We note here that the errors in $S'$ shown in Table \ref{table1} are computed taking different values of relative reflectance around 0.55 and 0.90 $\mu$m and checking how the values of S' change, i.e., the greater the dispersion of the data points in relative reflectance, the larger the error. However, we know from our observational experience, that division by the spectra of the solar analog introduces an intrinsic error in the spectral slope that is not smaller than 0.5\%/10$^3$\AA. This applies also to the near-infrared. Therefore, and to be as realistic as possible, we will use this value, unless the computed errors for S' are larger.

The observed difference in spectral slope could be caused by the difference in the solar phase angle between the two observations. While the data from Binzel et al. (\cite{binzel01}) was obtained in January 26, 1998, with a phase angle of $\sim$3.0$^{\circ}$, the visible spectrum presented in this work was taken at a phase angle of 18.0$^{\circ}$. This corresponds to a change in spectral slope of $\Delta$$S'$ = 0.15 $\pm$ 0.10 \%/10$^3$\AA/$^{\circ}$ for 3$^{\circ}$ $<$ $\alpha$ $<$ 18$^{\circ}$, which is in good agreement with the measured values of the phase reddening effect in the visible found in the literature: Lumme \& Bowell (\cite{lumme81}) measured $\Delta$$S'$ = 0.15  $\pm$ 0.17 \%/10$^3$\AA /$^{\circ}$ for a sample of C-types, while Luu \& Jewitt (\cite{luu90}) measured $\Delta$$S'$ = 0.18 \%/10$^3$\AA /$^{\circ}$ for 0$^{\circ}$ $<$ $\alpha$ $<$ 40$^{\circ}$ for a sample of near-Earth and main belt asteroids.

   \begin{figure}
   \centering
   \includegraphics[width=8cm]{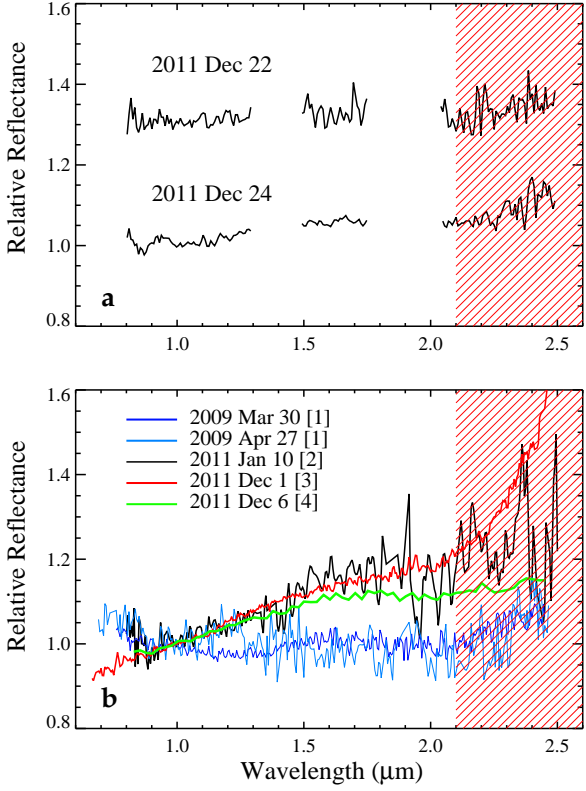}
      \caption{Near-infrared spectra of FG3. (a). Two new spectra presented in this work. Spectra are normalized to unity at 1.0 $\mu$m and offset for clarity. (b). Spectra of FG3 previously published: [1] MIT-UH-IRTF (MINUS, http://smass.mit.edu/minus.html); [2] de Le\'on et al. (2011); [3] Binzel et al. (2012); [4] Rivkin et al. (2012). The spectra are normalized to unity at 1.0 $\mu$m. We note the change in spectral slope. The red-dashed region is associated with thermal excess.}\label{figure2}
   \end{figure}

\subsection{Near-infrared spectra}

   \begin{figure}
   \centering
   \includegraphics[width=8cm]{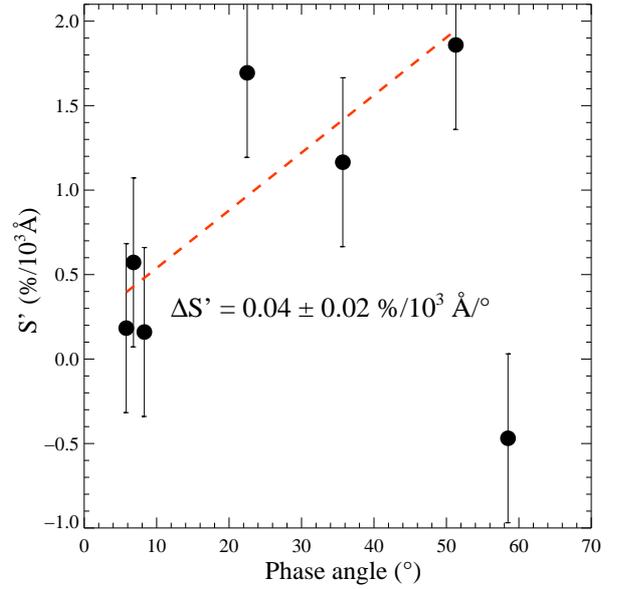}
      \caption{Spectral slope $S'$ computed for the near-infrated spectra {\it vs.} phase angle ($\alpha$). The represented values are listed in Table \ref{table1}. The red-dashed line corresponds to a linear fit to the points (excepting the point at the low-right part of the plot), with $\Delta$$S'$ being the slope of that linear fit.}\label{figure3}
   \end{figure}

In the case of the near-infrared, there are several published spectra of FG3. References, observational dates, distance to the Sun ($r$), and phase angle ($\alpha$) of these spectra are shown in Table \ref{table1}, and the spectra are plotted in Fig. \ref{figure2}b using different colors. From oldest to newest we have: two spectra from March and April 2009 (MIT-UH-IRTF, smass.mit.edu/minus.html), one spectrum from January 2011 (de Le\'on et al. \cite{deleon11}), and two spectra from December 2011 (Binzel et al. \cite{binzel12}, Rivkin et al. \cite{rivkin12}). The two near-infrared spectra presented in this work are also from December 2011 (see Fig. \ref{figure2}a). As in the case of visible wavelengths, we find a significant variation in the spectral slope. We compute it using the same definition, but normalizing at 1.0 $\mu$m (see also Table \ref{table1}). Most of the near-infrared spectra show an increase in reflectance beyond 2.1 $\mu$m, associated with thermal excess. The exception is the spectrum corresponding to December 6, 2011, which has been digitized from Rivkin et al. (\cite{rivkin12}), and that was already corrected from thermal excess. Therefore, to compute spectral slopes in a homogeneous way, we use the range between 0.9 and 2.1 $\mu$m. The computed values of $S'$ are shown in Table \ref{table1}. Figure \ref{figure3} shows $S'$ against phase angle. If one point at the low-right part of the plot is not considered, the data seem to show a trend. The computed variation, shown as a red-dashed line in the plot, is $\Delta$$S' = 0.04 \pm 0.08  \%/10^3\AA / ^{\circ}$ for $6^{\circ} < \alpha < 51^{\circ}$. We note that we have already taken into account the uncertainties in spectral slope introduced by the division by the solar analogs and described in Sect. 3.1, and we still see a trend in the data. The phase reddening effect has been mainly studied at visible wavelengths, but it has also been  observed in the near-infrared region. However, there are just a few references in the literature to properly compare with our obtained value. Nathues (\cite{nathues10}) analyzed the visible and near-infrared spectra of 97 asteroids belonging to the Eunomia collisional family, but they only provide values for $\Delta$$S'$ in the visible wavelength range. Clark et al. (\cite{clark02}) studied the near-infrared spectrometer observations (0.8 to 2.4 $\mu$m) of the S-type asteroid (433) Eros obtained by the NEAR Shoemaker spacecraft. They computed the spectral slope from 1.49 to 2.36 $\mu$m, finding a variation of $\Delta$$S'$ = 0.05 $\%$/10$^3$\AA / $^{\circ}$ for $0^{\circ} < \alpha < 100^{\circ}$.  Finally, Sanchez et al. (\cite{sanchez12}) studied the effects of phase reddening in the laboratory spectra of a sample of ordinary chondrites. They computed the spectra slope fitting a continuum across the 1 $\mu$m absorption band (between $\sim$0.7 and $\sim$1.55 $\mu$m), and found a variation ranging from $\Delta$$S'$ = 0.04 $\%$/10$^3$\AA / $^{\circ}$  for LL chondrites to $\Delta$$S'$ = 0.02 $\%$/10$^3$\AA / $^{\circ}$ for H chondrites for $13^{\circ} < \alpha < 120^{\circ}$.  They found that this effect is more intense for $\alpha >$ 30 $^{\circ}$. 
   
\section{Conclusions}\label{conc}
We have presented here three additional spectra of binary NEA 1996 FG$_3$, one in the visible and two in the near-infrared. Treating the two wavelengths separately and comparing with previous published spectra, we find in both cases a significant change in spectral slope. In the visible region, the change in spectral slope can be explained by the phase reddening effect, with a quantified variation that is in good agreement with the values found in the literature. In the case of the near-infrared spectra, we find a trend between the reddening of the spectra and the increase in phase angle, but only if we do not take into account one point. Unfortunately, we do not have access to the observational details of the spectrum corresponding to that point, and so we cannot check if there are any problems with it. Our computed variation in $S'$ is in good agreement with the only two values found in the literature for the phase reddening in the near-infared. Therefore, although we cannot firmly conclude it, it seems that the observed variation in the spectral slope in the near-infrared could also be explained by the phase reddening effect. In the case of binary asteroid FG3, one should take into account other factors, like the position of the secondary with respect to the primary, or the spin axis orientation at the time of the observations. More data needs to be collected and analyzed in order to properly explain the differences in spectral slope. 

\begin{acknowledgements}
      J.dL. acknowledges financial support from the current Spanish ``Secretar\'{\i}a de Estado de Investigaci\'on, Desarrollo e Innovaci\'on'' (Juan de la Cierva contract). J.L. and V.AL. acknowledge support from the projects AYA2011-29489-C03-02 and AYA2012-39115-C03-03 (MINECO). H.C. acknowledges support from NASA's NEOO program and from the National Science Foundation. This paper is based on observations made with the Italian Telescopio Nazionale Galileo (TNG) operated on the island of La Palma by the Centro Galileo Galilei of the INAF (Instituto Nazionale di Astrofisica) at the Spanish Observatorio del Roque de los Muchachos of the Instituto de Astrofisica de Canarias.
\end{acknowledgements}


\begin{thebibliography}{}

  \bibitem[2008]{barucci08} Barucci, M. A., Fornasier, S., Dotto, E., et al. 2008, A\&A, 477, 665
  \bibitem[2001]{binzel01} Binzel, R. P., Harris, A. W., Bus, S. J., \& Burbine, T. H. 2001, \icarus, 151, 139
  \bibitem[2004a]{binzel04a} Binzel, R. P., Rivkin, A. S., Stuart, J. S., et al. 2004a, \icarus, 170, 259
  \bibitem[2004b]{binzel04b} Binzel, R. P., Perozzi, E., Rivkin, A. S., et al. 2004b, Meteoritics and Planetary Science, 39, 351 
	\bibitem[2012]{binzel12} Binzel, R. P., Polishook, D., DeMeo, F., et al. 2012, Lunar and Planetary Institute Science Conference Abstracts, 43, 2222
  \bibitem[2010]{campins10} Campins, H., Morbidelli, A., Tsiganis, K., et al. 2010, \apj, 721, L53
  \bibitem[2003]{christou03} Christou, A. A. 2003, \planss, 51, 221
	\bibitem[2002]{clark02} Clark, B. E., Helfenstein, P., Bell III, J. F., et al. 2002, \icarus, 155, 189
  \bibitem[2010]{deleon10} de Le\'on, J., Licandro, J., Serra-Ricart, M., et al. 2010, \aap, 517, A23
	\bibitem[2011]{deleon11} de Le\'on, J., Moth\'e-Diniz, T., Licandro, J., et al. 2011, \aap, 530, L12
	\bibitem[1990]{jewitt90} Jewitt, D. C. \& Luu, J. X. 1990, \aj, 100, 933
  \bibitem[1992]{landolt92} Landolt, A. U. 1992, \aj, 104, 340 
	\bibitem[1981]{lumme81} Lumme K., \& Bowell, E. 1981, \aj, 86, 1705
	\bibitem[1990]{luu90} Luu, J. X. \ Jewitt, D. C. 1990, \aj, 99, 1985
  \bibitem[2000]{mottola00} Mottola, S., and Lahulla, F. 2000, \icarus, 146, 556
	\bibitem[2010]{nathues10} Natnues, A. 2010, \icarus, 208, 252
  \bibitem[2012]{popescu12} Popescu, M., Birlan, M. and Nedelcu, D. A., 2012, \aap, 544, 130 
  \bibitem[2000]{pravec00} Pravec, P., \v{S}arounov\'{a}, L., Rabinowitz, D. L., et al. 2000, \icarus, 146, 190
	\bibitem[2012]{rivkin12} Rivkin, A. S., Howell, E. S., DeMeo, F. E., et al. 2012, Lunar and Planetary Institute Conference Abstracts, 43, 1537
	Abstracts, 43, 1537
	\bibitem[2012]{sanchez12} Sanchez, J. A., Reddy, V., Nathues, A., et al. 2012, Icarus, 220, 36
  \bibitem[2009]{scheirich09} Scheirich, P., \& Pravec, P. 2009, \icarus, 200, 531
	\bibitem[2012]{walsh12} Walsh, K. J., Delb\'o, M., Mueller, M., et al. 2012, \apj, 748, 104
\end{thebibliography}
\end{document}